\documentstyle[12pt]{article}
\def\cp{{\it CP}}

\def\nbs{{n_B \over s}}

\def\chimns{\cmatrix{\psi_1\cr\psi_3}}
\def\chipls{\cmatrix{\psi_4\cr\psi_2}}
\def\cmatrix#1{\left(\matrix{#1}\right)}
\def\del{\Delta(\omega)}
\def\dt{\Delta t}
\def\dpo{\Delta p^0}
\def\dv{\vec\partial}
\def\dz{\partial_z}

\def\id{{\tt 1 \hskip -.3em l}}
\def\jz{j_z}
\def\m{{\cal M}}
\def\md{\m^{\dagger}}

\def\o#1{{\cal O}\left(#1\right)}
\def\psiz{\cmatrix{L\cr R\cr}}
\def\pz{-i\dz}
\def\sv{\vec\sigma}
\def\tr#1{{\rm Tr}\left[#1\right]}

\title{ Electroweak Baryogenesis and Standard Model $CP$ Violation }

\author{Patrick Huet and Eric Sather\thanks{ Work supported by the
   Department of Energy, contract DE-AC03-76SF00515. }\\
Stanford Linear Accelerator Center \\
Stanford University \\
Stanford, California 94309 \\ }
\date{}

\begin{document}

\maketitle

\vspace{-4in}
\rightline{ SLAC-PUB-6479 }
\rightline{ April 20, 1994 }
\rightline{ T/E/AS }
\vspace{ 3.2in}

\title{Abstract}

\begin{abstract}

\noindent

We analyze the mechanism of electroweak baryogenesis proposed by
Farrar and Shaposhnikov in which the phase of the CKM mixing matrix
is the only source of $CP$ violation.  This mechanism is based on a
phase separation of baryons via the scattering of quasiparticles by
the wall of an expanding bubble produced at the electroweak phase
transition. In agreement with the recent work of Gavela, Hern\'andez,
Orloff and P\`ene, we conclude that QCD damping effects reduce the
asymmetry produced to a negligible amount. We interpret the damping
as quantum decoherence. We compute the asymmetry analytically. Our
analysis reflects the observation that only a thin, outer layer of
the bubble contributes to the coherent scattering of the
quasiparticles. The generality of our arguments rules out any
mechanism of electroweak baryogenesis that does not make use of a
new source of $CP$ violation.

\end{abstract}

\vspace{ .3in }

\begin{center}{Submitted to: {\it Physical Review D} }\end{center}
\thispagestyle{empty}
\newpage

\setcounter{page}{1}

\section{Introduction}

The present work addresses the possibility of implementing the phase
of the CKM mixing matrix of the quarks as the source of $CP$
violation for electroweak baryogenesis.

The origin of the baryon asymmetry of the universe ({\it BAU}) is
recognized as a fundamental question of modern physics.  Although the
{\it BAU} is a macroscopic property of the entire observable
universe, the ingredients for its explanation are contained in the
microscopic laws of particle physics, as pointed out by Sakharov
\cite{Sakharov}.

Sakharov established on general grounds that a theory of particle
interactions could account for the production of the {\it BAU} at an
early epoch of the universe, provided that this theory contains
$B$-violating processes which operated in a $C$- and $CP$-violating
environment during a period when the universe was out of thermal
equilibrium.

The state of the art in particle physics is the Standard Model of
gauge interactions among quarks and leptons. $CP$ violation has been
observed and is thought to originate from the quark mixing matrix.
$B$ violation is believed to have taken place through
non-perturbative weak-interaction processes in the hot plasma of the
early universe.

Kuzmin, Rubakov and Shaposhnikov \cite{krs} pointed out that
implementing the program of Sakharov in the Standard Model would
require the electroweak phase transition to be first order, with the
baryon asymmetry being produced at the interface of bubbles of
nonzero Higgs expectation value, which expand into the unbroken
phase. Furthermore, Shaposhnikov \cite{bound} established a stringent
upper bound on the Higgs mass by requiring that the resulting baryon
asymmetry not be washed out by the $B$-violating processes from which it
originated. The latest studies \cite{dhs,phase} of the
electroweak phase transition have refined this bound to a value which
is now ruled out by experiment.  Although a better understanding of
the nonperturbative sector of the electroweak theory is required,
this bound directly challenges the possibility of electroweak
baryogenesis.

The above obstacle, however, is not the principal reason which has
motivated various groups to enlarge the framework of the Standard
Model in the search for a viable scenario of baryogenesis
\cite{review}. In the Standard Model, all $CP$ violation results from
a single complex phase in the quark mixing matrix.  This phase can be
transformed away in the limit that any two quarks of equal charge
have the same mass, and it can appear in physical observables only
through processes which mix all three generations of quarks. These
limitations suppress $CP$-violating effects in the Standard Model for
most processes by a factor of the order of $10^{-20}$. Given that
$CP$ violation is a necessary ingredient for baryogenesis, it is
difficult to reconcile this suppression factor with the observed
ratio of the baryons per photon in the Universe,
$(4$--$6)\times10^{-11}$.

Recently, Farrar and Shaposhnikov (FS) \cite{farrar} performed a
detailed analysis of this important question. Despite all
expectations, they concluded that Standard-Model $CP$ violation does
not lead to the above suppression; instead, they found that under
optimal conditions it is sufficient for generating a ratio of
baryons per photon of as much as the observed $10^{-11}$. A crucial
ingredient of their analysis is the interaction of the quarks with
thermal gauge and Higgs bosons in the plasma, which they correctly
take into account by expressing the interaction between the quarks
and the bubble interface as the scattering of quasiparticles.

Subsequently, Gavela, Hern\'andez, Orloff and P\`ene (GHOP) raised
objections to this analysis \cite{ghop}.  They pointed out that
Farrar and Shaposhnikov did not take into account the
quasiparticle width (damping rate).  The width results from
the fast QCD interactions of the quasiparticles with the plasma, and
is larger than any other scale relevant to the scattering. They
proposed to take the damping into account, and they concluded that it
reduces the magnitude of the {\it BAU} produced by the FS mechanism
to a negligible amount, in agreement with earlier expectations.  The
details of their analysis will appear in future publications.

We propose a novel interpretation of the damping rate, $\gamma$, of a
quasiparticle as a measure of its limited quantum coherence.  The
quasiparticle wave is rapidly damped because the components of the
wave are rapidly absorbed by the plasma, and reemitted in a different
region of the phase space. This decoherence phenomenon prevents
components of the wave from participating in quantum interference
over a distance longer than a {\it coherence length}, $\ell$, whose
magnitude is proportional to 1/$\gamma$. Quantum interference is
necessary for the generation of a $CP$-violating observable.

The above considerations lead us to reexamine the physical mechanism
of scattering of a particle off a medium. The latter does not take
place at the interface but instead results from the coherent
interference of components of the particle wavefunction which are
refracted by the bulk of the scattering medium. This observation can
be ignored if the incoming wave is coherent for an arbitrary amount
of time, but not for a quasiparticle which has a coherence
length much shorter than any other relevant scale. This perspective
provides a transparent physical understanding of the scattering
properties of a quasiparticle off the bubble.  The coherent
scattering of a quasiparticle effectively takes place only in a very
thin outer layer of the bubble, which drastically reduces the
probability of reflection.

In order to contribute to a $CP$-violating observable, a
quasiparticle wave must scatter many times in the bubble before it
decoheres.  It must encounter mixing of all three generations of
quarks and the $CP$-odd phase in the CKM matrix.  The scattering
takes place through the quark mass term in the bubble of broken
phase, and through interaction with charged Higgs in the plasma.
However, the mean free path for each of these scatterings is far
longer than the coherence length of the quasiparticle wave.  The wave
has almost completely died away by the time it has scattered a
sufficient number of times. Consequently, the baryon asymmetry
produced is insignificant, orders of magnitude smaller than the
observed asymmetry (and the asymmetry found by Farrar and
Shaposhnikov).

We make the above arguments quantitative by deriving a diagrammatic
expansion for the reflection of a quasiparticle wave off a bubble.
This expansion expresses a reflection amplitude as a sum of paths in
the bubble with various flavor changes and chirality flips, with each
path being damped by the exponential of its length expressed in units
of the coherence length $\ell$. This method provides an analytic
expression for the baryon asymmetry and demonstrates that the leading
order contributions are proportional to the Jarlskog determinant and
to an analogous invariant measure of $CP$ violation. Our analysis
corroborates the findings of GHOP that the $BAU$ produced is
suppressed to a negligible amount as result of plasma effects.

Our arguments of decoherence are of great generality and rule out
any scenario of baryogenesis which implements the phase of the CKM
matrix as the sole source of $CP$ violation.

In Section~2, we review the main aspects of the electroweak phase
transition which are needed to carry out our analysis and we describe
the FS mechanism of baryogenesis. In Section~3, we introduce and
justify the concept of the coherence length, and we describe the
physics of the scattering which takes into account the limited
coherence of the quasiparticles. Using these insights, we describe in
Section~4 our method for computing the baryon asymmetry in presence
of a sharp bubble wall.  We discuss various additional suppressions
which occur when the wall has a more realistic thickness.  Finally,
we summarize our results and discuss their applicability to more
general situations. In particular, we briefly discuss possible
implications for other scenarios of electroweak baryogenesis.

\section{ The Mechanism of Farrar and Shaposhnikov }

In this section, we review the relevant features of the electroweak
phase transition, and we describe the FS mechanism of electroweak
baryogenesis.

\subsection{ The Electroweak Phase Transition }

It is well established after the pioneering work of Kirzhnits and
Linde \cite{lindek} that the electroweak $SU(2)\times U(1)$ gauge
symmetry was unbroken in the early universe. As the universe cooled
down to a temperature of order $T\sim100\,$GeV, the thermal
expectation value of the Higgs field developed a nonzero value,
breaking the electroweak symmetry.

This phase transition is thought to have been a first-order
transition, although currently unresolved difficulties related to the
non-abelian gauge sector of the thermal plasma have prevented a proof
of this statement. Electroweak baryogenesis relies on this assumption
in order to meet the criteria of Sakharov.  In a second-order phase
transition, the departure from thermal equilibrium results from the
time dependence of the temperature, which is driven by the expansion
of the universe. The rate of expansion of the universe, $H=T^2/
M_{\rm Planck},$ is typically $17$ orders of magnitude slower than a
typical process in the plasma, far too slow to generate a significant
departure from equilibrium. On the other hand, in a first-order phase
transition the Higgs VEV jumps suddenly to a nonzero value. This
triggers the nucleation of bubbles of broken phase.  As a bubble
expands, its surface sweeps through the plasma, requiring a given
species to suddenly adjust its thermal distribution to its nonzero
mass inside the bubble.  This produces a temporary state of
nonequilibrium with a time scale of the order
(thickness)/(velocity)$\sim 10^{1\hbox{-}3}/T$, which is comparable
to the microscopic time scale of the plasma.

The dynamics of bubble expansion are fairly well understood. These
bubbles grow to a macroscopic size of order $10^{12}/T$ until they
fill up the universe. In contrast, baryogenesis is a microscopic
phenomenon $\sim(1$--$100)/T$. This allows one to ignore
complications due to the curvature of the wall by assuming the latter
to be planar. The thickness of the interface is of order
$(10$--$100)/T,$ depending on the Higgs mass, while the terminal
velocity of expansion $v_W$ has been calculated to be
non-relativistic \cite{phase,larryvel}, with the smallest allowed
velocity, of order $0.1,$ attained in the thin-wall limit.
Furthermore, for this range of parameters, the growth of the bubble
has been shown to be stable \cite{stablewall}.

The above considerations lead to a picture of the electroweak phase
transition favorable for the making of the baryon asymmetry.

\subsection{The Mechanism of Farrar and Shaposhnikov}

Farrar and Shaposhnikov proposed a simple mechanism of baryogenesis
based on the observation that as the wall sweeps through the plasma, it
encounters equal numbers of quarks and antiquarks which reflect
asymmetrically as a result of the $CP$-violating interactions
\cite{farrar}.  This mechanism leads to an excess of baryons inside
the bubble and an equal excess of antibaryons outside the bubble.
Ideally, the excess of baryons outside is eliminated by baryon
violating processes while the excess inside is left intact, leading
to a net {\it BAU}.

Outside the bubble is the domain of the unbroken phase. There are
rapid $B$-violating processes which occur at a rate per unit volume of
$\Gamma_{\rm out}= \kappa (\alpha_W T)^4$. The coefficient $\kappa$
is not reliably known, but Monte Carlo simulations \cite{montecarlo}
suggest $\kappa \sim .1$--$1$. These processes cause the baryon
asymmetry to relax to a thermally-averaged value of zero. A
fraction of the antibaryon excess escapes annihilation by diffusing
back inside the bubble, an effect enhanced by the motion of the wall,
and which can be accounted for by solving diffusion equations
\cite{farrar}.

Inside the bubble, the known $B$-violating processes are instanton
processes \cite{instanton}, which can be ignored because they occur
at a rate smaller than the expansion rate of the universe, and
sphaleron processes \cite{krs}, which occur at a rate $\Gamma_{\rm
in}\sim \exp(-2g_W\langle\phi\rangle/\alpha_W T)$. In order to
prevent the loss of the baryon excess in a subsequent epoch, the
latter processes must occur at a rate smaller than the expansion rate
of the universe: $\Gamma_{\rm in} \ll  (T^2/M_{\rm Planck})T^3$.
Since the expectation value $\langle\phi\rangle$ behaves
parametrically as $1/m_H^2$, this constraint yields an upper bound on
the Higgs mass \cite{bound,phase} of order $45\,$GeV, which lies
below the current experimental limit of $58\,$GeV \cite{limits}. This
conflict is a major difficulty for Standard-Model baryogenesis. It
can be resolved either by a drastic reformulation of sphaleron
physics or by extending the parameter space of the symmetry-breaking
sector. Both avenues are the subject of active investigation.

\subsection{Optimal Parameters}

The goal pursued by Farrar and Shaposhnikov is to use the
$CP$-violating phase of the quark mixing matrix as the only source of
$CP$ violation for the phase separation of baryons. To discuss this
aspect, it is useful to eliminate complications due to other aspects
of baryogenesis such as the physics of the $B$-violating processes
and the structure of the wall. If it turns out that the mechanism
works within this simplified framework, one can reconsider the
analysis within the full setting. In the following, we select ideal
conditions which not only simplify the analysis but also optimize the
generation of the baryon asymmetry and make no reference to transport
phenomena.

We choose the following values for the $B$-violating rates:
\begin{equation}
\Gamma_{\rm in}=0\ , \qquad \Gamma_{\rm out}= \infty\ .
\label{eq:idealrates}
\end{equation}
The first condition prevents the wash out of the asymmetry inside the
bubble. The second instantaneously eliminates the excess of
antibaryons directly outside the wall without reference to any
diffusion process. These conditions clearly maximize the asymmetry
and allow one to express it directly in terms of the velocity of the
wall and the reflection coefficients for the scattering of
quasiparticles off the bubble.

For the parameters of the wall, we choose
\begin{equation}
\delta_W=0\ , \qquad \qquad  v_W\sim 0.1\ .
\label{eq:idealwall}
\end{equation}
A wall of zero thickness enhances the quantum-mechanical aspects of
the scattering of fermions off the bubble. In fact, we will show how
various suppression factors develop as the wall thickness increases
from $2$--$3/T$ to the more realistic value $10$--$100/T$ quoted
earlier. The limit of small thickness was shown \cite{phase,larryvel}
to be the limit of maximal damping of the motion of the wall in the
plasma, a situation for which calculations are reliable and yield the
above value of $v_W$.

Finally, following FS, we assume that the scattering effectively
takes place in $1+1$ dimensions. This choice simplifies the
calculation greatly. Its justification relies on the observation that
the kinematics of the scattering only involves the component of the
momentum perpendicular to the wall. In addition, forward scattering
produces a maximal change of helicity of the fermion, which is
required to produce an asymmetry. Restoration of the 3-dimensional
phase space can only suppress the asymmetry further.

\subsection{ A Formula for $n_B/s$ }

Under the above assumptions, we can derive a simple expression for
the ``baryon-per-photon ratio,'' $n_B/s$.

In the rest frame of the wall, at any given instant there is an equal
amount of quarks and antiquarks striking the wall from either side.
As a result of $CP$ violation, quarks and antiquarks scatter
differently in the presence of the bubble, and become asymmetrically
distributed between the broken and unbroken phases. By assumption,
the baryon number outside the bubble is immediately eliminated,
leaving an equal but opposite baryon number inside the bubble.
Therefore the net baryon number produced is minus the thermal average
of the baryon number in the unbroken phase. The baryon number in the
unbroken phase is the sum of the excess due to baryons from the
unbroken phase ($u$) which reflect off the bubble back into the
unbroken phase, and the excess due to baryons transmitted from the
broken phase ($b$) into the unbroken phase. Hence the net baryon
number produced is given by
\begin{eqnarray}
n_B=-{1\over3} \biggl\{ &&\int {d\omega\over 2 \pi} n_L^u(\omega)
   {\rm Tr}\left[
   R_{LR}^{\dagger}R_{LR}- \bar R_{LR}^{\dagger}\bar R_{LR}
   \right]
   \qquad + \qquad (L \leftrightarrow R)
   \nonumber \\
  \ + &&\int {d\omega\over 2 \pi} n_L^b(\omega)
   {\rm Tr}\left[
   T_{LL}^{\dagger}T_{LL}- \bar T_{LL}^{\dagger}\bar T_{LL}
   \right]
   \qquad + \qquad  (L \leftrightarrow R) \biggr\}\ .
\label{eq:nbone}
\end{eqnarray}
The factor of $1/3$ is the baryon number of a quark.
The quantities $R$ and $T$ are matrices in flavor space that contain
the reflection and transmission coefficients.  For example,
$R_{LR}^{fi}$ is the coefficient of reflection for a left-handed quark
of initial flavor $i$ which reflects into a right-handed quark
(conserving angular momentum) of final flavor $f$.  $\bar R_{LR}$
corresponds to the $CP$-conjugate processes, that is, right-handed
antiquarks reflecting into left-handed antiquarks. $T_{LL}$ and $\bar
T_{LL}$ contain the transmission coefficients of the corresponding
particles approaching the bubble wall from the interior. Expression
(\ref{eq:nbone}) simplifies greatly after using unitarity,
$T_{LL}^{\dagger}T_{LL} + R_{LR}^{\dagger}R_{LR}=\id$, and $CPT$
invariance, $R_{RL}=\bar R_{LR}$:
\begin{equation}
n_B= {1\over3}\biggl\{\int{d\omega\over 2 \pi}
              \bigl(n_L^u(\omega)-n_R^u(\omega)\bigr)
-
              \int{d\omega\over 2 \pi}
              \bigl(n_L^b(\omega)-n_R^b(\omega)\bigr)
\biggr\} \times \Delta(\omega)\ ,
\label{eq:nBonebis}
\end{equation}
where $\Delta(\omega)={\rm Tr}[R_{RL}^{\dagger}R_{RL} -
R_{LR}^{\dagger} R_{LR}] = {\rm Tr}[\bar R_{LR}^{\dagger} \bar R_{LR}
- R_{LR}^{\dagger} R_{LR}]$. The distributions
$n^{u,\,b}_{R,\,L}(\omega)$ are Fermi-Dirac distributions boosted to
the wall frame:
\begin{equation}
n(\omega)=n_0(\gamma(\omega-\vec v_W \cdot \vec p))=
{1 \over e^{\gamma(\omega-\vec v_W \cdot\vec p)/T}+1}
\label{eq:fermidirac}.
\end{equation}

For zero wall velocity, all thermal distributions are identical in
the wall frame so that contributions to $n_B/s$ in
eq.~(\ref{eq:nBonebis}) from the broken and unbroken phases cancel
each other, as do contributions from the scattering of left- and
right-handed particles. The motion of the wall introduces the
nonequilibrium conditions required for the generation of the baryon
asymmetry. The leading contribution to $n_B/s$ thus appears at first
order in $v_W$. Expanding eq.~(\ref{eq:fermidirac}) in powers of
$v_W$, using the value $v_w=0.1,$ and dividing by the entropy
density, $s=2\pi^2 g^* T/45 \simeq 45 T$,\footnote{$g^*$ is the
number of massless degrees of freedom in the plasma $\sim 103$.} we
find the ``baryon-per-photon'' ratio produced to be
\begin{equation}
\nbs\simeq {10^{-3} \over T} \int{d\omega\over 2 \pi}
              n_0(\omega)(1-n_0(\omega))
              {(\vec p_L- \vec p_R)\cdot \hat v_W\over T}
              \times\del\ +\ {\cal O}(v_W^2) \, .
\label{eq:nBtwo}
\end{equation}
The whole calculation of the baryon asymmetry now reduces to the
determination of the left-right reflection asymmetry
$\del$.

The non-trivial structure of the phase space is contained in the
factor $(\vec p_L- \vec p_R)\cdot \hat v_W/T.$ This vanishes
unless, as discussed in the following subsection, interactions with
the $W$ and $Z$ bosons in the plasma are taken into account in the
propagation of the quarks; there we will see that $(\vec p_L-
\vec p_R)\cdot \hat v_W/T \sim \alpha_W$. In addition, the $CP$-odd
quantity $\Delta(\omega)$ vanishes unless flavor mixing interactions
occur in the process of scattering. This requires us to take into
account the interactions with the charged $W$ and Higgs bosons in the
scattering process. At first, this might appear an insurmountable
task. However, Farrar and Shaposhnikov suggested that all the
relevant plasma effects can consistently be taken into account by
describing the process as a scattering of suitably-defined
quasiparticles off the wall.

\subsection{ Quasiparticles }

Quasiparticles are fermionic collective excitations in a plasma.
They were studied decades ago in a relativistic context in an
$e^+$-$e^-$ plasma \cite{quasiabelian}. They were considered for the
first time in the QCD plasma by Klimov \cite{klimov} and Weldon
\cite{weldon1}. In the vacuum, a massless spin-1/2 particle with
energy $\omega$ and momentum $\vec p$ has the inverse propagator
$S_0^{-1} = \gamma^0\omega - \vec\gamma \cdot\vec p$. In the
plasma, the particle is dressed, acquiring a thermal self-energy
of the form
\begin{equation}
\Sigma(\omega, \vec p\,) = \gamma^0 a(\omega, p) -
b(\omega, p) \vec\gamma \cdot \vec p \ .
\end{equation}
The dispersion relations for the quasiparticles are obtained by
solving for the poles of the full propagator, including the self-energy.
We need to solve
\begin{equation}
\det[S_0^{-1} - \Sigma(\omega, \vec p\,)] = 0\ .
\label{eq:determinant}
\end{equation}
The solution is
\begin{equation}
\omega=a(\omega, p) \pm p [1-b(\omega, p)]\ .
\label{eq:solutiondet}
\end{equation}
The quantity $a(\omega,p)$ has a nonzero value, $\Omega,$ at zero
momentum, so that there is a mass gap in the dispersion relations. A
peculiar feature of this solution is the appearance of two branches
as seen in Fig.~1. The upper, ``normal'' branch ($n$) corresponds to
a ``dressed'' quark propagating as if it had an effective mass
$\Omega$. The second, ``abnormal'' branch ($a$) is interpreted
\cite{weldon2} as the propagation of a ``hole,'' that is the absence
of a antiquark of same chirality but opposite momentum. A ``hole'' is
expected to be unstable at large momentum, but is thought to be
stable for relatively small momentum \cite{smilga}, which is the
region of momentum of interest in the FS mechanism.

At small quasiparticle momentum, where the largest phase
separation of baryons occurs, the self-energy can be linearized as
\begin{equation}
\Sigma(\omega, \vec p) \simeq \gamma^0 (\Omega -\omega)
- \vec\gamma\cdot\vec p/3\ .
\label{eq:linselfenergy}
\end{equation}
The solutions for the poles in the quasiparticle propagator are in
this approximation simply
\begin{equation}
\omega \simeq \Omega \pm {p\over 3}\ .
\label{eq:lineardisprel}
\end{equation}
Here the factor of $1/3$ is the quasiparticle group velocity,
$d\omega/dp$, at zero momentum.

In the hot plasma of the early universe, left- and right-handed
quasiparticles acquire distinct thermal masses $\Omega_L$ and
$\Omega_R$ because only left-handed quarks couple to the thermal $W$
bosons.  The thermal masses also develop flavor dependence because
different flavors couple with different strength to the thermal Higgs
bosons. The thermal masses of the left-handed quasiparticles are
given explicitly by \cite{weldon1,farrar}
\begin{equation}
\Omega^2_L={2\pi\alpha_s T^2\over3}+
   {\pi\alpha_W T^2\over2}\left(
   {3\over4}+{\sin^2\theta_W\over36}+
   {M_u^2 + K M_d^2 K^{\dagger} \over 4 M_W^2} \right)\ ,
   \label{eq:omegal}
\end{equation}
where the contributions from thermal interactions with gluons,
electroweak gauge bosons, and Higgs bosons are all apparent.  In this
expression, $K$ is the CKM matrix, $M_u={\rm diag}(m_u, m_c, m_t),$
$M_d={\rm diag}(m_d, m_s, m_b)$, and the Yukawa couplings to the
Higgs have been related to the masses of the quarks
and the $W$ in the broken phase. For right-handed up quarks,
\begin{equation}
\Omega^2_R={2\pi\alpha_s T^2\over3}+
   {\pi\alpha_W T^2\over2}\left({4\sin^2\theta_W\over9}+
   {M_u^2 \over M_W^2}\right)\ ,
   \label{eq:upomegar}
\end{equation}
while for right-handed down quarks,
\begin{equation}
\Omega^2_R={2\pi\alpha_s T^2\over3}+
   {\pi\alpha_W T^2\over2}\left({\sin^2\theta_W\over9} +
   {M_d^2 \over M_W^2}\right)\ .
   \label{eq:downomegar}
\end{equation}
These results for the thermal masses hold at leading order in the
temperature, $T,$  assuming that $T$ is much larger than any other
energy scale.  In section~4, we will see that in order to have flavor
mixing of right-handed quarks, we need to consider corrections
proportional to $\log m/T$ that arise when nonzero quark masses in
the broken broken phase are taken into account.

The full structure of the dispersion relations (\ref{eq:solutiondet})
for left- and right-handed particles in the broken and the unbroken
phases is depicted in Fig.~2. The graphs which contribute to the
self-energy are of the form shown in Fig.~3a, where the quark
interacts with either a gluon, a $W$ boson or a Higgs boson in the
plasma.  The dominant contribution to the $\Omega$'s is left-right- and
flavor-symmetric, and comes from gluon exchange diagrams.  This is
contained in the left-right average of the $\Omega$'s which, ignoring
the small flavor-dependent pieces from Higgs and hypercharge-boson
interactions, is given by
\begin{equation}
\Omega_0 \simeq {g_s T \over\sqrt{6}}
   \left (1+{9\alpha_W \over 64 \alpha_s}\right)
   \simeq 50\,{\rm GeV}\, .
\label{eq:omega0}
\end{equation}
Splitting between left- and right-handed excitations comes dominantly
from the $W^{\pm}$ interactions,
\begin{equation}
\delta\Omega = {\Omega_L-\Omega_R}
   \simeq {g_W^2 T^2 \over 20 \Omega_0}
   \simeq 4\,{\rm GeV}\, .
\label{eq:omegadifference}
\end{equation}

In the unbroken phase, the energy levels of left- and right-handed
quasiparticles intersect at an energy close to $\Omega_0,$ at a
momentum $|\vec p|$ near $(3/2)\delta\Omega$. In the broken phase,
level-crossing takes place, leaving a mass gap of thickness equal to
the mass of the quark at the core of the quasiparticle. This is shown
in Fig.~2. Quasiparticles with such energies cannot propagate in the
broken phase; they are totally reflected by the bubble if they
approach it from the unbroken phase. This latter property is of
crucial importance in the FS mechanism and restricts the relevant
phase space to a region near $\omega = \Omega_0.$

Finally, there are other contributions to the self-energy resulting
from neutral- and charged-Higgs bosons. Their effects are unimportant
for the propagation of a quasiparticle in either phase.  However, the
self-energy contributions from interactions with the charged Higgs
are crucial for the generation of the baryon asymmetry. Without them,
the thermal masses would be flavor independent, and in a
mass-eigenstate flavor basis, the CKM matrix --- the only source of
$CP$ violation --- would not be present. With the charged-Higgs
interactions included, it is impossible to diagonalize the evolution
equations for the quasiparticles simultaneously in both phases: the
required $CP$-violating flavor mixing will be present in one or both
phases, allowing the separation of the baryons across the bubble wall.

\subsection{ Phase Separation of Baryon Number}

It is known that a $CP$-violating observable is obtained by
interfering a $CP$-odd phase, ${\cal B}$, with a $CP$-even phase,
${\cal A}$, so that, schematically, the asymmetry resulting from the
contribution of particles and antiparticles is proportional to
\begin{equation}
|{\cal A}+{\cal B}|^2-|{\cal A}+{\cal B}^*|^2=
-4 {\rm Im}\,{\cal A}\, {\rm Im}\,{\cal B}\ .
\label{eq:cpobservable}
\end{equation}
This illustrates the role of quantum mechanics in the generation of a
$CP$-odd observable. Farrar and Shaposhnikov proposed to describe
the scattering of quasiparticles as completely quantum mechanical,
that is, by solving the Dirac equation in the presence of a
space-dependent mass term. In particular, they identified the source
of the phase separation of baryon number as resulting from the
interference between a path where, say, an $s$-quark (quasiparticle)
is totally reflected by the bubble with a path where the $s$-quark
first passes through a sequence of flavor mixings before leaving the
bubble as an $s$-quark. The $CP$-odd phase from the CKM mixing matrix
encountered along the second path interferes with the $CP$-even phase
from the total reflection along the first path. Total reflection
occurs only in a small range of energy of width $m_s$ corresponding
to the mass gap for strange quarks in the broken phase, as depicted
in Fig.~2. This leads to a phase space suppression of order $m_s/T$.
Inserting this suppression into (\ref{eq:nBtwo}) yields the following
estimate of the Farrar and Shaposhnikov baryon-per-photon ratio:
\begin{eqnarray}
\nbs &\simeq&  10^{-3} \alpha_W {m_s \over T}
\bar\Delta \nonumber\\
     &\simeq& 10^{-7}\,\times\, {\bar \Delta} \, .
\label{eq:nBthree}
\end{eqnarray}
This estimate requires ${\bar \Delta}$, the energy-averaged value of
the reflection asymmetry, to be at least of order $10^{-4}$ in order
to account for the baryon asymmetry of the universe; this value is
just barely attained in Ref.~\cite{farrar}.

Gavela et al.\ pointed out that the above analysis ignores the
quasiparticle width, or damping rate, embodied by the imaginary part
of the self-energy
\begin{equation}
\Sigma={{\rm Re}\,\Sigma} - 2i \gamma .
\label{eq:gamma}
\end{equation}
The width results from the exchange of a gluon with a particle in the
plasma, and has been computed at zero momentum as $\gamma \simeq 0.15
g_s^2 T \simeq 20$\,GeV \cite{damping}. GHOP made the important
observation that this spread in energy is much larger than the mass
gap $\sim m$ in the broken phase, and as a result largely suppresses
$\del$. Their arguments rely on the analytic continuation in the
$\omega$-plane of the coefficients of reflection for quasiparticle
scattering.

In the next Section, we describe the role of the damping rate $\gamma$
in the scattering of a quasiparticle off the bubble from a perspective
which provides a clear physical understanding along with an unambiguous
computational method.

\section{ Coherence of the Quasiparticle }

\subsection{ The Coherence Length $\ell$ }

A Dirac equation describes the relativistic evolution of the
fundamental quarks and leptons. Its applicability to a quasiparticle
is reliable for extracting on-shell kinematic information, but one
should be cautious in using it to extract information on its
off-shell properties. A quasiparticle is a convenient bookkeeping
device for keeping track of the dominant properties of the
interactions between a fundamental particle and the plasma. For a
quark, these interactions are dominated by tree-level exchange of
gluons with the plasma. It is clear that these processes affect the
coherence of the wave function of a propagating quark. To illustrate
this point, let us consider two extreme situations.
\begin{itemize}
\item {\it The gluon interactions are infinitely fast}. In this case,
the phase of the propagating state is lost from point to point. A
correct description of the time evolution can be made in terms of a
totally incoherent density matrix. In particular, no interference
between different paths is possible because each of them is
physically identified by the plasma.\footnote{Not only is the quantum
mechanics of interference suppressed, but also the scattering
process is entirely classical.} As a result, no $CP$-violating
observable can be generated and $\Delta(\omega)=0$.
\item {\it The gluon interactions are extremely slow}. The
quasiparticle is just the quark itself and is adequately described by a
wavefunction solution of the Dirac equation, which corresponds to a
pure density matrix. In particular, distinct paths cannot be identified
by the plasma, as the latter is decoupled from the fermion. This
situation was implicitly assumed in the FS mechanism. This
assumption, however, is in conflict with the role the plasma plays in
the mechanism, which is to provide a left-right asymmetry as well as
the necessary mixing processes. In addition, this assumption is in
conflict with the use of gluon interactions to describe the kinematical
properties of the incoming (quasi-)quark.
\end{itemize}
The actual situation is of course in between the two limits above.
The quasiparticle retains a certain coherence while acquiring some of
its properties from the plasma. Whether this coherence is sufficient
for quantum mechanics to play its part in the making of a
$CP$-violating observable at the interface of the bubble is the
subject of the remaining discussion.

The damping rate $\gamma$ characterizes the degree of coherence of
the quasiparticle. It results from 2-to-2 processes of the type
shown in Fig.~3b. It is a measure of the spread in energy, $\Delta E
\sim 2\gamma,$ which results from the ``disturbance'' induced by the
gluon exchanged between the quark and the plasma. From the
energy-time uncertainty relation, $1/(2\gamma)$ is the maximum
duration of a quantum mechanical process before the quasiparticle is
scattered by the plasma.  We define a {\it coherence length} $\ell$ as
the distance the quasiparticle propagates during this time:
\begin{equation}
\ell = v_g\times {1 \over 2\gamma} \simeq {1\over 6\gamma}
\simeq {1 \over 120\,{\rm GeV} }\ ,
\label{eq:length}
\end{equation}
where $v_g$ is the group velocity of the quasiparticle. With this
definition, we can easily describe the decoherence that occurs during
the scattering off a bubble. Of crucial importance for the remaining
discussion, the coherence length of the quasiparticle is much shorter
than any other scale relevant to the scattering process:
\begin{equation}
\ell \simeq {1 \over T}\ \ll\
{1\over p}\simeq {1\over\ \delta\Omega}\simeq {20 \over T}\ , \
{1\over m_s} \simeq {1000 \over T}\ .
\label{eq:scales}
\end{equation}

\subsection{ A Model for Decoherence }

Having identified the limited coherence of a quasiparticle, we need
to describe its impact on the physics of scattering by a bubble of
broken phase. To understand this impact, let us first consider a
familiar example, the scattering of light by a refracting medium.

According to the  microscopic theory of reflection of light, the
refracting medium can  be decomposed into successive layers
of scatterers which diffract the incoming plane wave. The first layer
scatters the incoming wave as a diffracting grid. Each successive
layer reinforces the intensity of the diffracted wave and sharpens its
momentum  distribution. As more layers contribute to the
interference, the diffracted waves resemble more and more the full
transmitted and reflected waves. This occurs only if the wave
penetrates the wall {\it coherently} over a distance large compared
to its wavelength $k^{-1}$.

In analogy with the microscopic theory of reflection of light by a
medium, we can slice the bubble into successive layers which scatter
the incoming wave. The wavefunction for a quasiparticle reflected
from the bubble is the superposition of the waves reflected from each
of the layers.  However, the decoherence of the quasiparticles
arising from collisions with the plasma implies that quasiparticles
reflected from deep inside the bubble back into the symmetric phase
cannot contribute coherently to the reflected, outgoing wave of
quasiparticles.  Having traveled several coherence lengths through
the plasma, a component of the wave reflected from deep inside the
bubble will have been repeatedly absorbed and reemitted by the
plasma.  Each component thereby acquires a distinct momentum and
energy, preventing quantum interference of their amplitudes.
Therefore, scattering from layers of the bubble deeper than one
coherence length does not contribute significantly to the production
of a coherent outgoing wave.

We can make the above arguments more specific in three different but
complementary ways:
\begin{itemize}
\item The scattering occurs because of the gain in mass by the
quark when it enters the broken phase; this increment of mass is very
small, and the full scattering requires the {\it coherent}
contribution of scatterers up to a distance $1/m$ into the bubble in
order for the latter to probe the energy of the wave with a
resolution smaller than $m$. This requirement is not satisfied since,
from (\ref{eq:scales}), this minimal penetration length is 3 orders
of magnitude larger than the coherence length of the incoming wave.
\item From a corpuscular point of view, since scattering in the
bubble is due to the quark mass $m$, the mean free path for
scattering is $1/m.$  This is $1000$ times longer than the coherence
length.  Therefore the probability for quasiparticle scattering
even once in the bubble before it decoheres is extremely small, of
order $(m\ell)^2\sim 10^{-6}$.
\item Farrar and Shaposhnikov found a sizable baryon asymmetry
generated in an energy range of width $m_s$ where a strange quark is
{\it totally} reflected from the bubble. This energy range
corresponds to the mass gap in the broken phase described previously
(Fig.~2).  However, strange quarks can easily tunnel through a
barrier of thickness $\ell\ll1/m_s$, since they are off-shell by an
energy $\Delta\omega\simeq m_s$ for a time
$\dt\simeq\ell/v_g=1/(2\gamma)$. Because $\Delta\omega\dt
=m_s/(2\gamma)\ll1$, tunneling is completely unsuppressed and the
amplitude of the reflected strange-quark wave is only of order
$m_s/\gamma \sim 1/1000$.
\end{itemize}

The probability of scattering several times in the bubble, as is
required in order to generate a $CP$-violating, baryon asymmetry, is
thus vanishingly small. The baryon asymmetry results from
interference of reflected waves and necessarily involves several
flavor-changing scatterings inside the bubble in order to pick up the
$CP$-violating phase of the CKM matrix.  We therefore expect that the
baryon asymmetry produced when decoherence is properly taken into
account will be smaller than the amount found by Farrar and
Shaposhnikov by several factors of $m\ell$.

{}From these physical considerations, we can easily elaborate
quantitative methods of computing the scattering off a bubble by
quasiparticles with a finite coherence length $\ell$.

A simple model is obtained by expressing that when a quasiparticle
wave reaches a layer a distance $z$ into the bubble, its amplitude
will have effectively decreased by a factor $\exp(-z/2\ell).$   A
component which reflects from this layer and contributes to the
reflected wave will have decreased in amplitude by another factor of
$\exp(-z/2\ell)$ by the time it exits through the bubble wall.
We can take this into account by replacing the step-function
bubble profile with an exponentially decaying profile:
\begin{equation}
\hat M(z) = \cases{ Me^{-z/\ell},  & $z>0$\cr 0,  & $z<0$\cr}\ .
\end{equation}
This automatically attenuates the contribution to the reflected wave
from layers of the bubble deeper than one coherence length. The
analog in the theory of light scattering is the scattering of a light
ray by a soap bubble. For this reason, we refer to this model as the
``soap bubble" model. It is clear that truncating the bubble in this
way renders the bubble interface transparent to the quasiparticle, that
is, significantly reduces the amplitude of the reflected wave.

A more rigorous method of computing $\Delta(\omega)$ which we develop
in detail in the next Section is to solve an effective Dirac equation
in the presence of the bubble, including the decoherence (damping)
that results from the imaginary part of the quasiparticle
self-energy. We extract Green's functions which allows us to
construct all possible paths of the quasiparticles propagating in the
bulk of the bubble with chirality flips and flavor changes, each path
being damped by a factor $\exp(-{\cal L}/2\ell)$ where ${\cal L}$ is
the length of the path. Paths occurring within a layer of thickness
$\ell$ dominate the reflection amplitudes, in agreement with the
previous considerations. We refer to this method as the ``Green's
function" method.

We have computed $\Delta(\omega)$ using both methods. They give
results qualitatively and quantitatively in close agreement. The
principal difference is the following: The ``soap bubble" model
totally ignores scattering off the region deep inside the bubble, and
does not take into account small effects from decoherence in the
foremost layer. In the next Section we develop the ``Green's
function" method in detail.  The results are summarized in the final
Section.

\section{ Calculation of $\del$ Including Decoherence }

\subsection{ Dirac Equation for Quasiparticle Scattering }

In the unbroken phase where quarks are massless, quasiparticles
propagate with a well-defined chirality, and the wavefunctions $L$ and
$R$ of left- and right-handed quasiparticles evolve
independently according to
\begin{eqnarray}
(\omega + \vec\sigma \cdot\vec p - \Sigma_L(\omega, \vec p\,)) L
&=& 0\ ,  \nonumber\\
(\omega - \vec\sigma \cdot\vec p - \Sigma_R(\omega, \vec p\,)) R
&=& 0\ ,
\end{eqnarray}
where $\omega$ and $\vec p$ are the energy and momentum of the
quasiparticle, and $\Sigma_{L,\,R}$ are the thermal self-energies discussed
in Section~2.5.  The largest contribution to quasiparticle reflection and
the phase separation of baryons occurs at small momenta where the
momenta of left- and right-handed quasiparticles are not significantly
different.  At small momenta, the self-energies can be linearized
(\ref{eq:linselfenergy}) as $\Sigma_{L,\,R} \simeq  2(\Omega_{L,\,R} -
i\gamma) -\omega \pm\vec\sigma\cdot\vec p/3.$ Here
$\Omega_{L,\,R}$ are the thermal masses of left- and right-handed
quasiparticles introduced in eqs.~(\ref{eq:omegal}, \ref{eq:upomegar},
\ref{eq:downomegar}), and we have included the imaginary
damping term (\ref{eq:gamma}).

In the bubble of broken phase, the nonzero mass couples the two
chiralities of quasiparticles.  For an idealized bubble with a wall
of zero thickness at $z=0$ and extending to $z=+\infty,$ the mass
term is just $M\theta(z),$ where $M$ is the matrix of broken-phase
quark masses. The propagation and scattering of the quasiparticles in
the presence of the bubble of broken phase is thus governed by an
effective Dirac equation,
\begin{equation}
0=\cmatrix{2[\omega-\Omega_L+i\gamma + {1\over 3}
i\sv\cdot\dv] &M\theta(z)\cr
   M^\dagger\theta(z)&2[\omega-\Omega_R + i\gamma -
{1\over 3} i\sv\cdot\dv]\cr}
   \Psi(z) \ ,
\label{eq:lDirac}
\end{equation}
where $\Psi=\psiz.$ The field $\Psi$ can be either the field of the
down quarks, $(d,s,b)$, or the field of the up quarks $(u,c,t),$ and
in either case is a 3-component spinor in flavor space.  We ignore
the small corrections to this equation induced by boosting to the
frame of the bubble wall since they contribute at higher order in the
wall velocity. In a flavor basis which diagonalizes $\Omega_L$, this
Dirac equation is flavor-diagonal in the symmetric phase
($M\theta(z)=0$). Inside the bubble however,  flavors mix via the
mass matrix, which is off-diagonal in such a basis.  We treat the
mass matrix as a perturbation in order to make the calculation of the
quasiparticle reflection coefficients as physically transparent as
possible.  This is an excellent approximation for all quarks other
than the top, for which $m_t\ell\sim1.$  We will therefore
concentrate on the scattering of down quarks in this Section, and
describe qualitatively how these results would be altered for the
scattering of the top quark.  The large top mass does not alter the
implications of quasiparticle decoherence for the generation of a
baryon asymmetry.

Multiplying the above Dirac equation by $3/2,$ it becomes
\begin{equation}
\cmatrix{P_L + i\sv\cdot\dv&\m \theta(z) \cr
   \md \theta(z) &-(P_R + i\sv\cdot\dv)\cr}\Psi(z)=0\ ,
\label{eq:rDirac}
\end{equation}
where $P_L$ and $P_R$ are the symmetric-phase complex momenta of
the left- and right-handed quasiparticles, including the imaginary
damping terms,
\begin{eqnarray}
P_L &=&  3(\omega-\Omega_L+i\gamma)\\
P_R &=&- 3(\omega-\Omega_R+i\gamma)\ .
\label{eq:Pdefs}
\end{eqnarray}
The rescaled mass $\m$ is just given by $\m\equiv 3M/2.$
We can decompose $P_{L,\,R}$ into the physical (hermitian) momenta
$p_{L,\,R},$ and damping terms inversely proportional to the coherence
length $\ell=1/(6\gamma)$ introduced in eq.~(\ref{eq:length}):
\begin{eqnarray}
P_L = p_L - {i\over2\ell} &,&\
p_L = 3(\omega-\Omega_L)\ ;
\label{eq:pldef}\\
P_R = p_R + {i\over2\ell} &,&\
p_R = -3(\omega-\Omega_R)\ .
\label{eq:prdef}
\end{eqnarray}
The damping of the quasiparticle waves due to the imaginary
parts of $P_L$ and $P_R$ will be discussed shortly.

As discussed above, we restrict our attention to quasiparticles with
momenta perpendicular to the bubble wall.  Referring to the components of
$\Psi$ as
\begin{equation}
\Psi=\cmatrix{\psi_1\cr\psi_2\cr\psi_3\cr\psi_4\cr}\ ,
\end{equation}
we introduce spinors $\chi$ and $\tilde\chi$ for quasiparticles with
$j_z=\mp1/2$, where $j_z$ is the $z$-component of their angular
momentum:
\begin{equation}
\chi\equiv\chimns\ ; \tilde\chi\equiv\chipls\ .
\end{equation}
Because of angular momentum conservation, the Dirac
equation for $\Psi$ decomposes into two uncoupled equations, one for
$\jz=-1/2$ quasiparticles contained in $\chi$,
\begin{equation}
\pz\chi(z)=\cmatrix{P_L &\m\theta(z) \cr -\md\theta(z) &
P_R}\chi(z) \ ,
\label{eq:chimns}
\end{equation}
and another for the $\jz=+1/2$ quasiparticles contained in
$\tilde\chi$,
\begin{equation}
\pz\tilde\chi(z)=\cmatrix{-P_R &\md\theta(z) \cr -\m\theta(z) &
-P_L}\tilde\chi(z)\ .
\label{eq:chipls}
\end{equation}
In each of $\chi$ and $\tilde\chi$, the upper component represents a
quasiparticle moving towards the wall from the symmetric phase.   The
lower component represents a quasiparticle reflecting off the bubble back
into the symmetric phase. The $\jz=-1/2$ equation describes a left-handed
quasiparticle reflecting into a right-handed quasiparticle; the $\jz=+1/2$
equation describes the reversed process.

In the following we concentrate entirely on the scattering of
$\jz=-1/2$ quasiparticles contained in $\chi$.  To obtain analogous
results for the scattering of $\jz=+1/2$ quasiparticles, we need only
interchange $P_L\leftrightarrow -P_R$ and  $\m\leftrightarrow\md$, as
is apparent from eqs.~(\ref{eq:chimns}) and (\ref{eq:chipls}).

Consider the equation of motion for $\chi(z),$ eq.~(\ref{eq:chimns}),
keeping in mind the expressions for $P_{L,\,R}$ in
(\ref{eq:pldef}, \ref{eq:prdef}).
As stated above, $P_L$ and $P_R$ are the symmetric-phase momenta of
the left- and right-handed quasiparticles in $\chi.$   The signs of
the real parts of either $P_L$ or $P_R$ depend on whether
the quasiparticle is on the normal or abnormal branches, and this in
turn depends on the value of $\omega$ (see Fig.~1).  (For example, if
$\Omega_R<\omega<\Omega_L$, the left-handed quasiparticle is on the
abnormal branch and has negative momentum.  The right-handed
quasiparticle is in this case normal, but also has negative
momentum.) What is essential though, is that the sign of the group
velocities is independent of energy: the left-handed quasiparticles
move toward the bubble and positive $z$, and the
right-handed quasiparticles moves away from the bubble and in the
direction of negative $z$.

Now examine the imaginary parts of $P_L$ and $P_R.$  A left-handed
quasiparticle, which moves towards {\it positive} $z$, has a momentum
with a {\it positive} imaginary part.  Therefore the wavefunction for
left-handed quasiparticles decays as $\exp(-z/2\ell)$ as the
quasiparticles move towards positive $z$. A right-handed
quasiparticle, which moves toward {\it negative} $z$, has a momentum
with a {\it negative} imaginary part. Hence the wavefunction for
right-handed quasiparticles decays as $\exp(-|z|/2\ell)$ as the
quasiparticles move towards negative $z$. In other words, the
quasiparticles are damped no matter in which direction they propagate.

Note that we have implicitly chosen $\omega$ to be real. With
this choice, the momenta must become complex in order to satisfy the
dispersion relations, and propagation of quasiparticles in space is
damped. We have taken $\omega$ to be real because energy is conserved
in the scattering process.  We can then just ignore the factor
$\exp(-i\omega t)$ which describes the time dependence of the
quasiparticle wavefunction, since it does not affect the
probabilities of reflection.

We could have satisfied the dispersion relations with real momenta
if we had allowed $\omega$ to be complex.  Then we could have
observed the decay of the quasiparticles in a time $1/(2\gamma).$
But then the reflection probabilities would have an exponentially
decaying time dependence, which would require us to study the time
and space dependence of quasiparticle scattering in order to
determine the time it takes for a quasiparticle to scatter off the
bubble.

\subsection{Diagrammatic Calculation of Reflection Coefficients}

We now derive a perturbative expansion for the reflection
coefficients. The result is what one would intuitively expect: a
left-handed quasiparticle propagates toward positive $z$ until its
velocity is reversed by scattering in the bubble --- an insertion of
the quark-mass matrix --- and then becomes a right-handed
quasiparticle, propagating towards negative $z$, perhaps exiting the
bubble and contributing to the reflected quasiparticle wave, or
possibly scattering again, and once more propagating as a left-handed
deeper into the bubble. Throughout, the quasiparticle wave is damped.
To generate a  phase separation of baryons, the quasiparticle wave
must suffer a sufficient number of scatterings inside the bubble,
both with the neutral Higgs condensate, which gives factors of the
quark-mass matrix, and with charged Higgs in the plasma, in
order to produce a $CP$-violating observable.

First consider the propagation of quasiparticles in the symmetric
phase (again, restricting our attention to the $j_z=-1/2$
quasiparticles contained in $\chi$).   For the left- and right-handed
quasiparticles contained in $\chi$ we need to find Green's
functions $G_L$ and $G_R$ satisfying
\begin{equation}
(-i\dz - P_{L,\,R})G_{L,\,R}(z-z_0)=\id\,\delta(z-z_0)\ .
\end{equation}
In addition we require the boundary conditions
\begin{equation}
G_L(-\infty)=G_R(+\infty) = 0\ ,
\end{equation}
which state that there are no sources of quasiparticles at spatial
infinity. The unique solution is
\begin{eqnarray}
G_L(z-z_0) & = &\  i \theta(z-z_0) e^{iP_L(z-z_0)}
=\ i \theta(z-z_0) e^{-(z-z_0)/2\ell} e^{ip_L(z-z_0)} \ , \\
G_R(z-z_0) & = & -i \theta(z_0-z) e^{iP_R(z-z_0)}
= -i \theta(z_0-z) e^{-(z_0-z)/2\ell} e^{ip_R(z-z_0)} \ .
\end{eqnarray}
The $\theta$-functions indicate that left-handed quasiparticles move
toward positive $z$ while the right-handed quasiparticles move toward
negative $z$, as expected.  We have substituted the expressions for
$P_{L,\,R}$ (\ref{eq:pldef}, \ref{eq:prdef}) to demonstrate that
quasiparticle propagation is damped.

Now introduce the quark-mass terms in the bubble as a perturbation,
and consider the reflected wave of right-handed quasiparticles at $z=0$
due to a delta-function source of left-handed quasiparticles at $z=0.$
Let
\begin{equation}
\chi=\cmatrix{\chi_L \cr \chi_R}\ .
\end{equation}
We thus need to solve (\ref{eq:chimns})
\begin{eqnarray}
(-i\dz-P_L)\chi_L(z) &=& - i\delta(z)\chi_L(0) +
   \m\theta(z)\chi_R(z)\ , \\
(-i\dz-P_R)\chi_R(z) &= &-\md\theta(z)\chi_L(z)\ .
\end{eqnarray}
{}From the equations satisfied by the Green's functions we see that the
solution is given by
\begin{eqnarray}
\chi_L(z) & = & -i G_L(z)\chi_L(0) +
   \int dz_0 G_L(z-z_0) \m\theta(z_0) \chi_R(z_0)\ ,\\
\chi_R(z) & = & \int dz_0 G_R(z-z_0) (-\md) \theta(z_0) \chi_L(z_0)\ ,
\label{eq:chisolution}
\end{eqnarray}
where the integrals are over all $z_0$. These expressions can be
iterated to find the reflected wave to any order in the quark mass
matrix.

The reflection matrix $R_{LR}$, where the subscript indicates that
left-handed quasiparticles are reflected into right-handed
quasiparticles, is obtained by considering all possible flavors of
initial and final quasiparticles.  For example, $R_{LR}^{fi}$, the
reflection coefficient for scattering of initial flavor $i$ into a
final flavor $f$, is found by calculating the $f$-component of
$\chi_R(0)$ when the $i$-component of $\chi_L(0)$ is set equal to
one and the other components are set to zero.  From the solution
eq.~(\ref{eq:chisolution}), we see that the reflection matrix is
given by the expansion
\newpage
\begin{eqnarray}
&R_{LR}&= -i \int dz_1G_R(-z_1)(-\md)\theta(z_1)G_L(z_1)
\nonumber \\
&-&i\int dz_1 dz_2 dz_3 G_R(-z_3)(-\md)\theta(z_3)G_L(z_3-z_2)
\m \theta(z_2) G_R(z_2-z_1)(-\md)\theta(z_1)G_L(z_1)
\nonumber\\
&&+ \cdots \\
&=& i\int_0^\infty dz_1 e^{-iP_R z_1} \md e^{iP_L z_1}
\nonumber\\
&+&i\int_0^\infty dz_1 \int_{z_1}^0 dz_2 \int_{z_2}^\infty dz_3\
e^{-iP_R z_3} \md e^{iP_L(z_3-z_2)} \m e^{iP_R(z_2-z_1)} \md e^{iP_L z_1}
+ \cdots\ .
\label{eq:Rexpansion}
\end{eqnarray}
This expansion is shown diagrammatically in Fig.~4.
 %
\setcounter{figure}{3}
\begin{figure}
\begin{center}
\begin{picture}(350,100)(-40,0)
\thicklines

\put(-80,40){\large$R_{LR}=$}

\put(-40,0){\line(1,0){40}} \put(0,0){\line(0,1){80}}
\put(0,80){\line(1,0){100}}

\thinlines

\put(-20,65){\vector(4,-1){60}}\put(40,50){\line(4,-1){40}}
   \put(80,40){\circle*{3}}\put(85,38){${\cal M}$}
\put(80,40){\vector(-4,-1){40}}\put(40,30){\line(-4,-1){60}}
\put(20,59){$L$} \put(20,14){$R$}

\put(130,40){\large$+$}

\thicklines

\put(156,0){\line(1,0){40}} \put(196,0){\line(0,1){80}}
\put(196,80){\line(1,0){100}}

\thinlines

\put(176,76){\vector(4,-1){52}}\put(228,63){\line(4,-1){28}}
   \put(256,56){\circle*{3}}\put(261,54){${\cal M}$}
\put(256,56){\vector(-4,-1){20}}\put(236,51){\line(-4,-1){20}}
   \put(216,46){\circle*{3}}\put(199,46){${\cal M}^\dagger$}
\put(216,46){\vector(4,-1){32}}\put(248,38){\line(4,-1){32}}
   \put(280,30){\circle*{3}}\put(285,28){${\cal M}$}
\put(280,30){\vector(-4,-1){40}}\put(240,20){\line(-4,-1){64}}

\put(220,68){$L$} \put(224,51){$R$}
\put(252,39){$L$} \put(220,5){$R$}

\put(330,40){\large$+\ \cdots$}

\end{picture}
\end{center}
\caption{First two terms in the expansion for the reflection matrix
$R_{LR}.$  The bubble of broken phase is indicated by the step. An
incident left-handed quasiparticle approaches the bubble from the
left, and is scattered by the quark-mass term $\m$ in the bubble,
becoming a right-handed quasiparticle which moves back towards the
bubble wall. The right-handed particle can then exit the bubble and
contribute to the reflected wave, or else can scatter again, via
$\md,$ leading to a contribution to the reflected wave at higher
order in the quark mass matrix.  The full reflected wave is obtained
by summing up these diagrams and integrating over the positions of
the scatterings in the bubble.}

\end{figure}
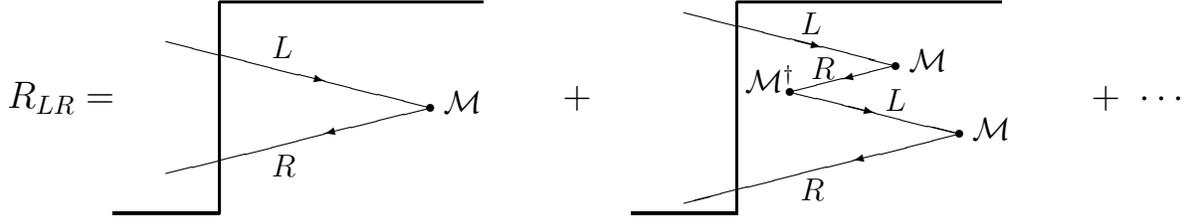

Let us now make explicit the damping of the quasiparticle waves.
Decomposing each of $P_L$ and $P_R$ into a momentum and a damping
term as in eqs.~(\ref{eq:pldef}, \ref{eq:prdef}), the previous
expression for $R_{LR}$ becomes
\newpage
\begin{eqnarray}
R_{LR} &=& i\int_0^\infty dz_1 e^{-z_1/\ell}e^{-ip_R z_1} \md
   e^{ip_L z_1}
\nonumber\\
&+&i\int_0^\infty dz_1 \int_{z_1}^0 dz_2 \int_{z_2}^\infty dz_3
\exp[-(z_1+|z_2-z_1|+|z_3-z_2|+z_3)/(2\ell)]\nonumber\\
&\times&
e^{-ip_R z_3} \md e^{ip_L(z_3-z_2)} \m e^{ip_R(z_2-z_1)} \md e^{ip_L z_1}
+ \cdots\ .
\label{eq:Rdamped}
\end{eqnarray}
The quasiparticle wave is evidently damped along each leg of its
trajectory.  The overall suppression for each term in $R_{LR}$ is
just $\exp(-{\cal L}/2\ell),$ where ${\cal L}$ is the distance
traveled by a quasiparticle in the barrier. In particular, it is
apparent that there will be no significant contribution to $R_{LR}$
from paths which travel to a depth of more than one coherence length
into the bubble.  Hence only an extremely thin outer layer of the
bubble contributes to the coherent reflected wave.

This perturbative expansion for the reflection matrix $R$ is the
basis for the calculations we are about to describe.  We will work
throughout to lowest nonvanishing order in the quark mass matrix.
This expansion is valid as long as $M\ell\ll1.$  This condition is
easily satisfied for all quarks other than the top, for which the
expansion parameter is of order unity, and for which our results will
only be qualitative.

We also work to lowest order in the ${\cal O}(\alpha_W)$
flavor-dependent terms in $p_{L,\,R}$ that arise from Higgs
contributions to the thermal self-energy. Decompose $p_{L,\,R}$ as
\begin{eqnarray}
p_L = p^0_L\id + \delta p_L\ ,\\
p_R = p^0_R\id + \delta p_R\ ,
\label{eq:dr}
\end{eqnarray}
where $p^0_{L,\,R}$ contain the large, flavor-independent terms in
$p_{L,\,R},$ while $\delta p_{L,\,R}$ are proportional to the
$\o{\alpha_W}$, flavor-dependent pieces in $\Omega^2_{L,\,R}$ that
arise from interactions with Higgs. Examining the expression
(\ref{eq:omegal}) for $\Omega^2_L,$ and formula (\ref{eq:pldef})
for $p_L,$ we see that for down quarks, $\delta p_L$ is given in a
mass-eigenstate flavor basis by
\begin{equation}
\delta p_L \simeq -{3\pi\alpha_W T^2\over16\Omega_0}
 {K^{\dagger} M_u^2 K + M_d^2 \over M_W^2}\ .
\label{eq:dlexpr}
\end{equation}
For the scattering of up quarks, $M_d$ and $M_u$ are interchanged, as
are $K$ and $ K^{\dagger}.$  The non-commutativity of $\delta p_L$
with $\m$ gives rise to flavor mixing in the broken phase.

The thermal masses $\Omega^2_R$ for right-handed quarks are
flavor-diagonal when approximated for large temperatures, $T$
(\ref{eq:upomegar}, \ref{eq:downomegar}).  Hence in this
approximation, $p_R$ is diagonal and does not contribute to the
flavor mixing required for $CP$ violation. In the broken phase, the
quarks appearing in the self-energy graphs are massive,
and $\Omega_R^2$ acquires off-diagonal terms (which are usually
neglected at large temperatures).  As pointed out by GHOP, the
resulting off-diagonal terms in $\delta p_R$, which do not commute
with the quark mass matrix, lead to additional contributions to
$\Delta(\omega)$. For down quarks,
\begin{equation}
\delta p_R = \cdots + {3\alpha_W\over32\pi\Omega_0}
 {M_d K^\dagger\, M_u^2\log(M_u^2/T^2)\, K M_d \over M_W^2}+\cdots\ ,
\label{eq:drexpr}
\end{equation}
where we have omitted the flavor-independent terms of order $T^2.$
All masses are high-temperature, broken-phase masses.  Again, for the
scattering of up quarks, $M_d\leftrightarrow M_u$ and
$K\leftrightarrow K^\dagger.$

In addition to working to lowest nonvanishing order in the quark-mass
matrix $\m,$ we also work to lowest order in $\delta p_{L,\,R}$,
which is equivalent to lowest nonvanishing order in $\alpha_W.$
Given that in the range of momentum where our analysis is applicable
the diagonal components of $p_{L,\,R}$ are much smaller than
$1/\ell$, we could justifiably work to lowest order in $p^0_{L,\,R}$
as well.  However, we list results valid to all orders
in $p^0_{L,\,R}$ in order to show the energy dependence of
$\Delta(\omega)$.

We find two leading contributions to $\del$.  The first contribution
is the dominant contribution when quark masses are neglected when
calculating the self-energy in the broken phase. In this case $\delta
p_R$ in eq.~(\ref{eq:dr}) is diagonal and commutes with the quark
mass matrix. This is the only contribution considered by FS, and
comparing our results with FS we can see the dramatic effect of
decoherence. In a second calculation, we calculate the contribution
to $\Delta(\omega)$ that comes from including the off-diagonal terms
in $\delta p_R$.  GHOP found the largest contribution to
$\Delta(\omega)$ when considering the scattering of up-type quarks
with these terms included. In each case we take into account the
finite coherence length of the quasiparticle by using expansion
(\ref{eq:Rexpansion}) for the coefficient of reflection.

\subsection{Calculation of $\del$ Neglecting $\delta p_R$}

The leading contribution to $\del$ when $\delta p_R$ is ignored
appears at $\o{\m^6}$. In this case the momentum of the right-handed
quasiparticles is diagonal and commutes with the quark-mass matrix.
Then the expression for $R_{LR},$  eq.~(\ref{eq:Rexpansion}), can be
written as
\begin{eqnarray}
R_{LR}&=& i\int_0^\infty dz_1 \m e^{-z_1/\lambda}
\nonumber\\
&+&i\int_0^\infty dz_1 \int_{z_1}^0 dz_2 \int_{z_2}^\infty dz_3\
\m e^{-(z_3-z_2)/\lambda} \m^2 e^{-z_1/\lambda}
+ \cdots\ ,
\end{eqnarray}
where $1/\lambda=1/\ell-i(p_L-p_R^0).$ For simplicity, we have chosen
a mass-eigenstate basis where $\m$ is diagonal. Evaluating the
integrals, we find that
\begin{equation}
R_{LR}=i\m\lambda(\id - \m^2\lambda^2
+ \m^2\lambda^2\m^2\lambda^2+\m^2\lambda\m^2\lambda^3+\cdots)\ .
\end{equation}

To calculate $\Delta(\omega)={\rm Tr}(\bar R_{LR}^\dagger \bar R_{LR}
-R_{LR}^\dagger R_{LR}),$\footnote{The derivation of
eq.~(\ref{eq:nBonebis}) for $\Delta(\omega)$ made use of unitarity to
relate probabilities of reflection and transmission.  With damping,
unitarity might seem to violated.  However, the damping corresponds
to decoherence. Baryon number is still conserved throughout the
scattering process.} we need the reflection matrix $\bar R_{LR}$ for
the scattering of $CP$-conjugate particles.  The $CP$-conjugate
process differs only in that the CKM mixing matrix $K$ is replaced
with $K^*,$ which means $p_L$ is replaced with $p_L^*=p_L^T.$ Hence
the reflection matrix $\bar R_{L,\,R}$ is obtained from $R_{L,\,R}$
by replacing $\lambda$ with its transpose.  The ${\cal O}(\m^2)$ and
${\cal O}(\m^4)$ terms cancel out of the difference in
$\Delta(\omega),$ so that the leading-order contribution is
$\o{\m^6}$:
\begin{eqnarray}
\Delta_9(\omega)&=&{\rm Tr}[
\lambda\lambda^{\dagger 2} \m^2\lambda^\dagger \m^2 \lambda^{\dagger 2}
   \m^2 +
\lambda^2\lambda^\dagger   \m^2\lambda^2 \m^2 \lambda \m^2 +
\lambda^2\lambda^{\dagger 2} \m^2\lambda \m^2 \lambda^\dagger
\nonumber\\ &&-
\lambda\lambda^{\dagger 2} \m^2\lambda^{\dagger2} \m^2 \lambda^\dagger
   \m^2 -
\lambda^2\lambda^\dagger   \m^2\lambda \m^2 \lambda^2 \m^2 -
\lambda^2\lambda^{\dagger 2} \m^2\lambda^\dagger \m^2 \lambda]\ ,
\end{eqnarray}
where $1/\lambda^\dagger=1/\ell+i(p_L-p_R^0).$  Notice that each
factor of the quark mass matrix is accompanied by a factor of
$\lambda\simeq\ell.$  The product $\m\ell$ is the amplitude for
the quasiparticle to scatter through the quark mass term while
propagating for one coherence length, which is quite small.
To lowest order in $\delta p_L$,
\begin{eqnarray}
\Delta_9(\omega)&=&4i f_9(\Delta p^0\ell)
\,\tr{ (\delta p_L)^2 \m^4 \delta p_L \m^2 -
   (\delta p_L)^2 \m^2 \delta p_L \m^4 } \ell^9\\
&=&-{4i\over3} f_9(\Delta p_0\ell)\, {\rm Tr}
\left[\m^2, \ \delta p_L\right]^3 \ell^9\ ,
\label{eq:d9}
\end{eqnarray}
where $\dpo\equiv p_L^0-p_R^0,$ and $f_9$ is an energy-dependent form
factor given by
\begin{equation}
f_9(x) = {1\over (1+x^2)^6}\ .
\end{equation}
The subscripts ``9'' indicates that this contribution to
$\Delta(\omega)$ occurs at $9^{\rm th}$ order in $\ell.$  It is the
largest contribution to $\Delta(\omega)$ due to down quarks when
$\delta p_R$ is neglected. Again note that for every factor of the
quark mass matrix or $\delta p_L,$ there is an accompanying factor of
$\ell.$  Scattering from either the Higgs condensate or the the
charged Higgs in the plasma during one coherence length has a very
small probability.

Refering to the diagrammatic expansion in Fig.~4, this
${\cal O}(\m^6)$ contribution to $\Delta(\omega)$ evidently can come
from the interference of two paths which each have 3 chirality flips
(via the mass term), or it can come from the interference of a path
which has just one chirality flip with a path that has 5 chirality
flips.  The 3 factors of $\delta p_L$ are distributed among the
left-handed segments of the two paths.

We now substitute expression (\ref{eq:dlexpr}) for $\delta p_L$
and also $\Delta p^0\ell\equiv p_L^0-p_R^0 = (\omega-\Omega_0)/\gamma,$
using eqs.~(\ref{eq:pldef}) and (\ref{eq:prdef}), and where
$\Omega_0\simeq50\,$GeV is the left-right average of the
flavor-independent pieces of $\Omega_{L,\,R}$ introduced in
eq.~(\ref{eq:omega0}).  Our expression for $\Delta_9(\omega)$ for
down quarks becomes
\begin{equation}
\Delta_9^d(\omega)=-4\left({27\pi\alpha_W T^2\over 64\Omega_0
M_W^2}\right)^3  \Biggl[1+\Biggl(
{\omega-\Omega_0 \over \gamma}\Biggl)^2\Biggr]^{-6}
{\det{\cal C}\> \ell^9}\ ,
\label{eq:d9simple}
\end{equation}
The quantity $\det{\cal C}$ is the basis-independent Jarlskog
determinant \cite{Jarlskog},
\begin{eqnarray}
\det{\cal C} &=& i\det[M_u^2, K M_d^2 K^{\dagger}]\nonumber\\
&=&-2 J (m_t^2-m_c^2)(m_t^2-m_u^2)(m_c^2-m_u^2)
(m_b^2-m_s^2)(m_b^2-m_d^2)(m_s^2-m_d^2)\ ,
\label{eq:cquantity}
\end{eqnarray}
where the superscript $d$ indicates that this is the contribution to
$\Delta_9(\omega)$ due to the scattering of down quarks. Here $J$ is
the product of CKM angles $J=s_1^2s_2s_3c_1c_2c_3\sin\delta \sim
10^{-5}.$ Clearly, the largest contribution to $\Delta^d_9(\omega)$
comes from paths involving bottom quarks (either incident, reflected
or virtual).

For the scattering of up quarks our expansion in the quark-mass
matrix breaks down because of the large mass of the top quark.
Because of its large mass, the top quark is far off shell in the
broken phase (by $m_t-\Omega_0\simeq3\gamma$). We therefore expect
that if we did not treat the top mass as a perturbation, the
contributions from paths involving the top quark would be smaller
than the results obtained here.   Our results for the up quarks are
thus qualitative, and overestimate their contribution to the
asymmetry relative to the contribution of the down quarks.

As mentioned above, results for the scattering of up quarks can be
obtained from down-quark results by interchanging $M_d$ with $M_u$
and $K$ with $K^{\dagger}.$  From the definition of the Jarlskog
determinant in eq.~(\ref{eq:cquantity}), we see that it changes sign
under these interchanges. Hence to lowest order in $\m$, the
contribution to $\Delta_9(\omega)$ from up quarks,
$\Delta_9^u(\omega),$ differs only by a sign from the down-quark
contribution, $\Delta_9^d(\omega)$.  If the top were as light as the
other quarks, the total contribution to $\Delta_9(\omega)$ would
vanish (continuing to ignore the off-diagonal terms in $\delta p_R$).
Because the top is very heavy the dominant terms in
$\Delta^u_9(\omega),$ which come from paths involving top quarks,
will be reduced. Therefore, the total contribution to $\Delta_9$,
given by $\Delta_9^d+\Delta_9^u$, will not vanish.

We reserve further discussion of this contribution for the final
Section, and now describe our calculation of the leading contribution
to $\Delta(\omega)$ when the off-diagonal terms in $\delta p_R$ are
considered.

\subsection{Calculation of $\Delta(\omega)$ Including $\delta
p_R$}

Because $\delta p_R$ contains two factors of $M,$ when $\delta p_R$
is included, we need two fewer factors of the quark-mass matrix in
order to form an invariant analogous to the Jarlskog determinant.
The leading-order term therefore appears at ${\cal O}(M^4)$.

To find $\Delta(\omega)$ when the off-diagonal terms in $\delta p_R$
in eq.~(\ref{eq:drexpr}) are included, we again use the expansion for
$R_{LR}$ in eq.~(\ref{eq:Rexpansion}).  We can no longer directly
evaluate the $z$-integrals because of the noncommutativity of $\delta
p_L$ and $\delta p_R$ with $\m$.  Instead we first expand the integral
expression for $\Delta(\omega)$ in powers of $\delta p_L$ and $\delta
p_R$, and pick out the lowest-order nonvanishing terms, of order
$(\delta p_L)(\delta p_R)$. It is then possible to evaluate the
flavor-independent integral coefficient.  The resulting contribution
to $\Delta(\omega)$, at $7^{\rm th}$ order in $\ell,$ is
\begin{equation}
\Delta_7(\omega)=-8 i\,f_7(\dpo\ell)\,
\tr{\delta p_L\m\delta p_R\m^3 - \delta p_L\m^3\delta p_R\m}\ell^6\ ,
\label{eq:d7}
\end{equation}
where $f_7(\dpo\ell)$ is an energy-dependent form factor,
\begin{equation}
f_7(x) = {x \over (1+x^2)^4}\ .
\end{equation}
Note that $\Delta_7(\omega)$ would vanish if either $\delta p_L$ or
$\delta p_R$ commuted with $\m.$ Unlike $\Delta_9(\omega),$
$\Delta_7(\omega)$ is an {\it odd} function of $\dpo$ and so vanishes
at $\dpo=0$. This is because in order to discern the $CP$-odd phase
in the CKM matrix, we need a $CP$-even phase, as is apparent in
eq.~(\ref{eq:cpobservable}). Examining eq.~(\ref{eq:Rdamped}) for
$R_{LR}$, the only sources of relative phases are the factors of the
form $\exp(ipz).$  To get a nontrivial $CP$-even phase, we evidently
need an odd number of factors of $ip$. While the trace in
$\Delta_9(\omega)$ contains three $\delta p$'s, the trace in
$\Delta_7(\omega)$ contains just two, so we need a factor of
$\dpo$ to have a nontrivial $CP$-even phase.

Because $\Delta_7(\omega)$ is an odd function of $\dpo$, it vanishes
at $\omega=\Omega_0,$ in the middle of the energy range where light
quarks are totally reflected.  This is where Farrar and Shaposhnikov
saw the generation of a large baryon asymmetry, and where on a much
smaller scale, $\Delta_9(\omega)$ is peaked. We expect that
contributions to the integrated asymmetry from $\omega<\Omega_0$ will
largely cancel against contributions from $\omega>\Omega_0$, and
leaving a very small contribution to the integrated asymmetry from
$\Delta_7(\omega)$ for light quarks.

This contribution to the asymmetry arises from the interference of a
path that has three chirality flips with a path having one chirality
flip. The factor of $\delta p_L$ can occur along any of the
left-handed segments of the two paths, and similarly the factor of
$\delta p_R$ can occur along any of the right-handed segments.

Substituting the expressions for $\delta p_{L,\,R}$ for down quarks
in eqs.~(\ref{eq:dlexpr}, \ref{eq:drexpr}), and substituting $\Delta
p^0\ell= (\omega-\Omega_0)/\gamma$ as before, expression
(\ref{eq:d7}) for $\Delta_7(\omega)$ simplifies to
\begin{equation}
\Delta_7^d(\omega) = 2\left({27\alpha_W T \over 32 \Omega_0
M_W^2}\right)^2 f_7\left({\omega-\Omega_0 \over \gamma}\right)
{\cal D}_d \> \ell^6\ .
\label{eq:d7simple}
\end{equation}
The superscripts $d$ again indicates that this contribution is due to
the scattering of down quarks. The quantity ${\cal D}_d$ is an
invariant measure of $CP$ violation analogous to the Jarlskog
determinant:
\begin{eqnarray}
{\cal D}_d & = & {\rm Im\ Tr} \left[M_u^2\log M_u^2 \, K \, M_d^4 \,
  K^\dagger \,M_u^2 \, K \, M_d^2 \, K^\dagger \right]\nonumber\\
  &=&J \left[m_t^2 m_c^2 \log {m_t^2\over m_c^2} +
             m_t^2 m_u^2 \log {m_u^2\over m_t^2} +
             m_c^2 m_u^2 \log {m_c^2\over m_u^2} \right]\nonumber\\
&& \ \times  (m_b^2-m_s^2)(m_b^2-m_d^2)(m_s^2-m_d^2)\ .
\label{eq:dquantity}
\end{eqnarray}
Here we have used ${\rm Im}(K_{\alpha j} K^{\dagger}_{j\beta}
K_{\beta k} K^\dagger_{k\alpha})=J\sum_{\gamma,l}
\epsilon_{\alpha\beta\gamma} \epsilon_{jkl}$ \cite{Jarlskog}. Like
the Jarlskog determinant, ${\cal D}_d$ vanishes if any two quarks of
equal charge have the same mass.

Recall that the Jarlskog determinant (\ref{eq:cquantity}) simply
changes sign under the simultaneous interchanges $M_d\leftrightarrow
M_u$ and $K\leftrightarrow K^\dagger.$  By contrast, ${\cal D}_d$
does not treat the up-quark and down-quark mass matrices
symmetrically, and becomes a new quantity, ${\cal D}_u,$ under these
interchanges. This new quantity contains two more powers of $m_t,$
and is roughly $-1000$ times ${\cal D}_d.$ Hence the contribution to
$\Delta_7(\omega)$ due to up-quark scattering, $\Delta_7^u,$
obtained from the down-quark contribution by replacing ${\cal D}_d$
with ${\cal D}_u,$ will be much larger than $\Delta_7^d$.  Given that
contributions from paths including top quarks should be reduced when
the off-shellness of the tops is taken into account, the value for
$\Delta_7^u$ obtained here serves as an upper bound for  $\Delta_7.$

We now discuss our results for $\Delta(\omega)$ and their
implications for the size of the baryon asymmetry.

\section{ Presentation and Discussion of the Results }

\subsection{ Results }

In the previous Section we computed the energy-dependent reflection
asymmetry $\Delta(\omega)$. This asymmetry is the difference of
$\tr{\bar R_{LR}^\dagger \bar R_{LR}}$ and $\tr{R_{LR}^\dagger
R_{LR}}$, the probabilities for a left-handed quark and its
$CP$-conjugate to be reflected off the bubble, summed over all quark
flavors.  We calculated the reflection probabilities by solving an
effective Dirac equation including all relevant plasma effects as
self-energy corrections, in the presence of the space-dependent
mass term.

The real part of the self-energy accounts for the gluon interactions
which control the kinematical properties of the quarks. It accounts
for the interactions with the $W$'s which differentiate between
quarks with different chiralities, as well as interactions with the
charged Higgs which provide the flavor-changing processes needed for
the generation of a $CP$-violating observable. These effects are
embodied in the concept of quasiparticles which was used in the
mechanism of Farrar and Shaposhnikov.

The novelty of our calculation resides in our treatment of the
imaginary part of the self-energy. We interpreted the latter as a
measure of the coherence of the wave function of the quasiparticle,
and we introduced the concept of the coherence length, $\ell.$
We extracted Green's functions which, in conjunction with chirality
flips due to the mass term and flavor changes due to interactions with
the charged Higgs, lead to the construction of all possible paths
contributing to the reflection coefficients (Fig.~4). It is the
interference between these paths which survives in the asymmetry,
as expected from the general principles described in Sections~2 and
3. An important feature is that each path has an amplitude
proportional to exp$(-{\cal L}/2\ell),$ where ${\cal L}$ is the
length of the path. This confines the scattering to a layer of
thickness $\ell$ at the surface of the bubble, a property already
predicted on physical grounds in Section~3. The asymmetry results
from processes which involve a sufficient number of changes of flavor
as well as a sufficient number of factors of the quark mass matrix,
every one of which brings along a factor of $\ell$. Consequently, the
asymmetry is suppressed by many powers of $M\ell,$ the dimensionless
product of the quark-mass matrix and the coherence length of the
quasiparticle, $\ell,$ and powers of $(\delta p_L)\ell$ and $(\delta
p_R)\ell$, products of the coherence length with the flavor-dependent
terms in the momentum matrices for left- and right-handed
quasiparticles.

Specifically, we find the asymmetry dominated by:
(i) Contributions at order $\ell^7$ from processes involving the
scattering of up quarks, with two flavor
mixings, proportional to $(\delta p_L)(\delta p_R),$ and given in
eq.~(\ref{eq:d7simple}):\footnote{For our numerical results we set
$T=100\,$GeV.  We take the broken-phase $W$ mass as $M_W=T/2$, and
scale the broken-phase quark masses accordingly.  We use a generous
value for the product of sines and cosines of CKM angles:
$J=5\times10^{-5}.$}
\begin{eqnarray}
\Delta_7^u(\omega)
&=& -16\; f_7\Bigl(6\ell(\omega-\Omega_0)\Bigr)\; {\rm Im\ Tr }
[\delta p_L\m \delta p_R \m^3]
\; \ell^9
\label{eq:7trace}\\
&=& 2\left({27\alpha_W T \over 32\Omega_0 M_W^2}\right)^2 \,\times\,
f_7\Bigl(6\ell(\omega-\Omega_0)\Bigr)\,\times\, {\cal D}_u\>
\ell^6
\label{eq:dssimple} \\
&=& 10^{-18}\; f_7\Bigl(6\ell(\omega-\Omega_0)\Bigr)\ ,
\nonumber
\end{eqnarray}
where
\begin{equation}
f_7(x)=  { x \over (1+x^2)^4}\ ;
\label{eq:fseven}
\end{equation}
and (ii) Contributions at order $\ell^9$ from processes involving the
scattering of down quarks, with three flavor mixings, proportional to
$(\delta p_L)^3$, and given in eq.~(\ref{eq:d9simple}):
\begin{eqnarray}
\Delta_9^d(\omega)
&=&-8\; f_9\Bigl(6\ell(\omega-\Omega_0)\Bigr)\> {\rm Im\ Tr }
[(\delta p_L)^2 \m^4\delta p_L \m^2]\,\,\ell^9
\label{eq:9trace} \\
&=&-4\left({27\pi\alpha_W T^2\over 64\Omega_0 M_W^2}\right)^3
\,\times\, f_9\Bigl(6\ell(\omega-\Omega_0)\Bigr)\,\times\,
{\det{\cal C}\; \ell^9}
\label{eq:dnsimple} \\
&=& 4\times 10^{-22}\> f_9\Bigl(6\ell(\omega-\Omega_0)\Bigr)\ ,
\nonumber
\end{eqnarray}
where
\begin{equation}
f_9(x)=  { 1 \over (1+x^2)^6}\, .
\label{eq:fnine}
\end{equation}
The contribution to $\Delta_7(\omega)$ from down quarks is
$\simeq10^{-21}f_7\Bigl(6\ell(\omega-\Omega_0)\Bigr),$  while the
contribution to $\Delta_9(\omega)$ from up quarks is smaller than
the down-quark contribution.

Our results for up quarks should be regarded as upper bounds. In the
broken phase, the kinematics of the top quark is determined entirely
by its large mass, as opposed to the light quarks, whose kinematical
properties are dominated by their interaction with the plasma in both
phases. The reflection asymmetry is produced in an energy range near
where level-crossing occurs, well below the top quark mass. At these
energies the top quark can only propagate far off-shell. As discussed
in Section~4, this diminishes the amplitude for any path which
involves flavor changing from or to the top quark. In consequence,
the up-quark contribution to $\Delta_9(\omega)$ is suppressed
relative to the down-quark contribution, and the up quark
contribution to $\Delta_7(\omega)$ given in eqs. (\ref{eq:7trace})
and (\ref{eq:dssimple}) is an upper bound.

The two contributions $\Delta^u_7(\omega)$ and $\Delta^d_9(\omega)$
decompose naturally into a product of three factors, as given in
eqs.~(\ref{eq:dssimple}, \ref{eq:dnsimple}), each of which reflects
an important aspect of the physics involved.  Let us consider them
separately.

The first factor contains powers of $\alpha_W/M_W^2,$  which
originate from the flavor changing insertions $\delta p_L$ or $\delta
p_R$ on the path of the scattered quasiparticle.

The second factor is an energy-dependent function $f(x)$. Although,
the precise form of this function is sensitive to the details of the
calculation, its general shape is not. This function is a form factor
which reflects the increased likelihood of chirality flips at energies
for which the various flavors involved have similar momenta. That
occurs in the region of level crossing around $\omega \sim
\Omega_0\simeq 50\,$GeV (Fig.~2). The form factor peaks at a value of
order one, and have a width of order the quasiparticle width,
$\gamma.$  These properties are apparent in Fig.~5. Note that $f_9$
is peaked at $\omega=\Omega_0,$ while $f_7,$ though centered about
the same energy, actually vanishes there as the result of the
vanishing of the $CP$-even phase at that energy, as described in
Section~4.4.

Finally, the third terms on the right-hand sides of
eqs.~(\ref{eq:dssimple}) and (\ref{eq:dnsimple}) are the Jarlskog
determinant $\det{\cal C}$ and another $CP$-violating invariant, ${\cal
D}_u,$ which are given explicitly in eqs.~(\ref{eq:cquantity}) and
(\ref{eq:dquantity}) respectively. They contain the expected
dependence on the flavor mixing angles and vanish in the limit where
any two quarks with the same charge have equal masses.  We have
already argued that in general a \cp-violating observable such as
$\Delta(\omega)$ is the result of quantum interference between
amplitudes with different $CP$-even and $CP$-odd phases. These
physical processes can easily be identified from the structure of the
traces in eq.~(\ref{eq:7trace}) and (\ref{eq:9trace}). To do so, we
represent each of these traces as a closed fermion path with various
mass insertions ($M_d$) and flavor changing insertions ($\delta p_L$
and $\delta p_R$) in the order they appear in the trace. The mass
operator changes the chirality of the quark but not its flavor while
the flavor changing operator leaves the chirality intact. Any cut
performed across two portions of the loop with opposite chirality,
divides the loop into two open paths whose interference contributes
to the asymmetry. This is illustrated in Fig.~6. These paths are in
one-to-one correspondence with the paths constructed with the
Green's functions method elaborated in Section~4.

We now calculate the contributions to $n_B/s$. The contribution from
$\Delta_9^d(\omega)$ is, from eq.~(\ref{eq:nBthree}),
\begin{eqnarray}
\nbs\Biggl|_9&\simeq& {10^{-3}}\,\alpha_W\,{1\over T}
\int{d\omega\over 2 \pi}
              n_0(\omega)(1-n_0(\omega))\>
              \Delta_9(\omega) \nonumber\\
    &\simeq& { 10^{-25}}\,{1\over T} \int{d\omega\over 2 \pi}
              n_0(\omega)(1-n_0(\omega))\
              f_9\bigl(6\ell(\omega-\Omega_0)\bigr) \\
    &\simeq&  2 \times 10^{-28} \ .
\label{eq:nBresultnine}
\end{eqnarray}
Similarly, for the contribution from $\Delta_7^u(\omega)$
we find\footnote{The corresponding contribution to $n_B/s|_7$ from
down quarks is only $10^{-29}.$  Hence the largest
contribution to $n_B/s$ from the scattering of down quarks comes
from at ${\cal O}(\ell^9).$
For up quarks the ${\cal O}(\ell^7)$ contribution to $n_B/s$
is larger than the ${\cal O}(\ell^9)$ contribution.}
\begin{equation}
\nbs\Biggl|_7 \simeq -6 \times 10^{-27} \ .
\label{eq:nBresultseven}
\end{equation}
Because of the peculiarities of top quark kinematics, we cannot say
whether the up quark contribution to $n_B/s$ is in fact larger than
the contribution from down quarks given in
eq.~(\ref{eq:nBresultnine}). We therefore quote the result
(\ref{eq:nBresultseven}) as an upper bound on the magnitude of the
integrated asymmetry:
\begin{equation}
\Biggl|{n_B \over s}\Biggr| < 6 \times 10^{-27} \ .
\end{equation}

In Section~3, we advertised another method of computing the
asymmetry using a model in which the essentially infinitely thick
bubble is replaced with a thin layer of thickness $\ell$. We referred
to this model as the ``soap bubble" model. This model implements
quantum decoherence in scattering in the simplest way and provides an
analytic form of the asymmetry which has exactly the same structure
as the ones obtained in eq.~(\ref{eq:dssimple}) and
(\ref{eq:dnsimple}).  In fact, the only difference relative to the
results for $\Delta(\omega)$ obtained via the ``Green's function"
method is that for the ``soap bubble" model, the energy-dependent
form factors $f_7$ and $f_9$ are replaced with form factors $\hat
f_7(x)$ and $\hat f_9(x)$, where
\begin{eqnarray}
\hat f_7(x)&=& {2 x\over 3}
\>{1-{1\over 243}(23x^2+7x^4-3x^6)
\over (1+x^2)^4(1+(x/3)^2)^3 } \ ,
\\
\hat f_9(x)=&=&{1\over54}
\Biggl\{ {1-{1\over3}x^2 \over [(1+x^2)(1+(x/3)^2)]^3 }
+ \cdots \Biggr\}\ .
\end{eqnarray}
The terms omitted in $\hat f_9(x)$ are of order $1\%$ of the term
listed. These form factors differ slightly in form and magnitude from
their counterparts obtained via the ``Green's function" method, but
have the same overall shape.  For example, $\hat
f_9(6\ell(\omega-\Omega_0))$ is peaked at $\omega=\Omega_0$, while
$\hat f_7(6\ell(\omega-\Omega_0))$ vanishes at that energy. This
model leads to a baryon-per-photon ratio comparable in magnitude to
the values found in eqs.~(\ref{eq:nBresultnine}) and
(\ref{eq:nBresultseven}). We do not present the calculations for this
model in order to avoid redundancy.

Our results ought to be compared to the results of Farrar and
Shaposhnikov.  They calculated $\Delta(\omega)$ without taking into
account quasiparticle decoherence. They found a significant baryon
asymmetry  $n_B/s$ of order $10^{-11}$ from a region of energy for
which the strange quark is totally reflected. Taking into account the
decoherence of the quasiparticles, we find such total reflection to
be impossible and the asymmetry to be reduced to a negligible amount.
This conclusion corroborates the findings of Gavela,
Hern\'andez, Orloff and P\`ene (GHOP) \cite{ghop}.

Finally, we would like to comment on the more realistic situation of
quasiparticles interacting with a wall of nonzero thickness.
Typically, in the Standard Model and in most of its extensions, the
wall thickness $\delta$ is of order $10$-$100/T$, much larger than
the coherence length $\ell\sim 1/T$ and other typical mean free
paths.\footnote{Although, according to the authors of ref.
\cite{review}, the possibility of a thin wall is not ruled out.} As
the wall thickness increases from $0$ to a value a few times $\ell$,
the increment of mass over the latter distance is reduced by a factor
$\ell / \delta$ which has the effect to reducing the reflection
probabilities accordingly as a power law. As the thickness increases
further to a distance of a few wavelengths $k^{-1}\sim 5/T$, a WKB
suppression of order $\exp(-k\delta)$ is expected to turn on and to
suppress the process further. Clearly, the interior of a thick wall
is not a suitable environment for the occurrence of the subtle
quantum-mechanical phenomena which are to take place in order to
generate a $CP$-violating observable.

\subsection{Conclusions}

We have demonstrated that the FS mechanism operating at the
electroweak phase transition cannot account for the baryon asymmetry
of the universe. Our conclusions agree with the results obtained in
Ref.~\cite{ghop}.

Our arguments are powerful enough to establish more generally that
the complex phase allowed in the CKM mixing matrix cannot be the
source of $CP$ violation needed by any mechanism of electroweak
baryogenesis in the minimal Standard Model or any of its extensions.
Indeed, the generation of a $CP$-odd observable requires the quantum
interference of amplitudes with different $CP$-odd and $CP$-even
properties and whose coherence persists over a time of at least
$1/m_q$.  On the other hand, QCD interactions restrict the coherence
time to be at most $\ell \sim 1/(g_s^2 T)$, typically three orders
of magnitude too small. It is clear from the interpretation of the
Jarlskog determinant or any other $CP$- violating invariant we gave
in Section~5.1 and Fig.~6, that the processes necessarily proceed
through interference between amplitudes with multiple flavor mixings
and chirality flips; as a result, the asymmetry between quarks and
antiquarks appears to be strongly suppressed by many powers of $\ell
m_q$. This line of argument does not rely on the details of the
mechanism considered and can be applied to rule out any scenario of
electroweak baryogenesis which relies on the phase of the CKM matrix
as the only source of $CP$ violation.

QCD decoherence might be avoided in mechanisms which do not
involve light quarks. For example, the effect of decoherence is
negligible for the top quark: $\ell m_t \simeq 1$. A mechanism which
involves the scattering of only the top quark is viable, but at the
cost of introducing a new source of $CP$ violation \cite{ckntop}.
Other scenarios based on various extensions of the minimal standard
model such as the two-Higgs doublets \cite{lsvt} or SUSY \cite{dhss}
are also negligibly affected by the above considerations.

Although the Standard Model contains all three ingredients
required by Sakharov, it proves to be too narrow a framework
for an explanation of the baryon asymmetry of our universe.

\section*{ Acknowledgements }

We are grateful to Michael Peskin for helping us formulate our
picture of decoherence.  We thank Helen Quinn and Marvin Weinstein
for helpful discussions.

 \def\thefiglist#1{\section*{Figure Captions \markboth
    {FIGURE CAPTIONS}{FIGURE CAPTIONS}}\list
    {Figure \arabic{enumi}.}
    {\settowidth\labelwidth{Figure #1.}\leftmargin\labelwidth
    \advance\leftmargin\labelsep
    \usecounter{enumi}\parsep 0pt \itemsep 3pt plus 1pt}
    \def\newblock{\hskip .11em plus .33em minus -.07em}
    \sloppy}
\let\endthefiglist=\endlist
\begin{thefiglist}{9}
\item Schematic picture of the dispersion relations for a fermionic
quasiparticle in a hot plasma.  The upper curve represents the normal
branch.  The lower curve represents the abnormal branch, which
corresponds to the propagation of a ``hole.'' The abnormal branch
becomes completely unstable when it passes through the light cone
$\omega=p$ (dotted line) \cite{smilga}.

\item Dispersion curves linearized for small momentum $p$. Because
the $W$ and $Z$ bosons in the plasma only interact with left-handed
quasiparticles, the dispersion relations for left- and right-handed
quasiparticles are distinct. For a given chirality, the dispersion
relations are as shown in Fig.~1, with both a normal branch and an
abnormal branch.  (These curves are only shown for a single, light
flavor.  The curves for other light flavors would be shifted slightly
in energy.) In the unbroken phase, the left-handed abnormal branch
intersects the right-handed normal branch;  in the broken phase, the
nonzero quark mass connects the two chiralities and level crossing
occurs, as indicated by the dashed lines (here illustrated for the
charm quark). The result is a mass gap with thickness of order the
quark mass \cite{farrar}.

\item a) Graph contributing to the real part of the quasiparticle
self-energy. The dashed lines represent either gluons, electroweak
gauge bosons, or Higgs bosons from the plasma. These graphs are
responsible for the thermal masses $\Omega$ of the quasiparticles,
shown in Fig.~1.
 b)  Graph describing a collision of a quasiparticle (solid line)
with a quark or gluon (dashed line) in the plasma.  This graph
contributes to the imaginary part of the self-energy, and leads to
the decoherence of quasiparticle waves.

\item First two terms in the expansion for the reflection
matrix $R_{LR}.$  The bubble of broken phase is indicated by the
step. An incident left-handed quasiparticle approaches the bubble
from the left, and is scattered by the quark-mass term $\m$ in the
bubble, becoming a right-handed quasiparticle which moves back
towards the bubble wall. The right-handed particle can then exit the
bubble and contribute to the reflected wave, or else can scatter
again, via $\md,$ leading to a contribution to the reflected wave at
higher order in the quark mass matrix.  The full reflected wave is
obtained by summing up these diagrams and integrating over the
positions of the scatterings in the bubble.

\item The energy-dependent form factors $f_7$ and $f_9,$
evaluated at $(\omega-\Omega_0)/\gamma=6\ell(\omega-\Omega_0).$  Note
that $f_9$ is peaked at $\omega=\Omega_0\simeq50\,$GeV, while $f_7$
vanishes there.

\item This loop summarizes all the contributions to
$\Delta_9(\omega),$ and corresponds to the trace in
eq.~(\ref{eq:9trace}).  (An analogous loop summarizes the
contributions to $\Delta_7(\omega).$) The solid blobs represent
insertions of $\delta p_L,$ which describes the mixing of quark
flavors through interaction with the charged Higgs bosons.  The
crosses stand for insertions of the quark mass matrix.  The loop is
then a trace in flavor space of the product of all the insertions.
Any individual contribution to the reflection asymmetry can obtained
by cutting across two segments of the loop of opposite chirality.
This produces two open paths whose interference contributes to the
asymmetry, as shown in the right side of the figure.

\end{thefiglist}


\begin{thebibliography}{9}
 %
 %
\def\prdabbr{{\it Phys.\ Rev.}\  D}
\def\plbabbr{{\it Phys.\ Lett.}\ B}
\def\npbabbr{{\it Nucl.\ Phys.}\ B}
\def\prlabbr{{\it Phys.\ Rev.\ Lett.}}
 %
\def\prd#1#2#3{\prdabbr\ #1\ (19#2)\ #3}
\def\plb#1#2#3{\plbabbr\ #1\ (19#2)\ #3}
\def\npb#1#2#3{\npbabbr\ #1\ (19#2)\ #3}
\def\prl#1#2#3{\prlabbr\ #1\ (19#2)\ #3}
 %
\bibitem{Sakharov} A.~D.~Sakharov {\it JETP Lett.} {\bf 5} (1967) 24.
\bibitem{krs} V.~A.~Kuzmin, V.~A.~Rubakov and M.~E.~Shaposhnikov,
   \plb{155}{85}{36}.
\bibitem{bound} M.~E.~Shaposhnikov, \npb{287}{87}{757}.
\bibitem{dhs} M.~Dine, P.~Huet and R.~Singleton,~Jr.,
   \npb{375}{92}{625}.
\bibitem{phase} M.~Dine, P.~Huet, R. G.~Leigh, A.~Linde and D.
   Linde, \plb{283}{92}{319}; \prd{46}{92}{550}.
\bibitem{review} For a review of various attempts see A.~G.~Cohen,
   D.~B.~Kaplan and A.~E.~ Nelson, {\it Ann. Rev. Nucl. Part. Sci.},
   {\bf 43} (1993) 27. See also refs.~\cite{ckntop,lsvt,dhss}.
\bibitem{farrar} G.~R.~Farrar and M.~E.~Shaposhnikov, \prl{70}{93}{2833};
   preprint CERN-TH-6734/93, RU-93-11.
\bibitem{ghop} M.~B.~Gavela, P.~Hern\'andez, J.~Orloff and O.~P\`ene,
   preprint 1993, CERN 93/7081, LPTHE Orsay-93/48, HUTP-93/48,
   HD-THEP-93-45.
\bibitem{lindek} D.~A.~Kirzhnits and A.~D.~Linde, \plb{72}{72}{471}.
\bibitem{larryvel} L.~McLerran, B.-H.~Liu and N.~Turok,
   \prd{46}{92}{2668}.
\bibitem{stablewall} P.~Huet, K.~Kajantie, R.~G.~Leigh, L. McLerran and
   B.-H. Liu, \prd{48}{92}{2477}.
\bibitem{montecarlo} See for example J.~Ambjorn, T.~Askgaard, H.~Porter
   and M.~E.~Shaposhnikov, \plb{244}{90}{479}; \npb{353}{91}{346}.
\bibitem{instanton} A. A. Belavin, A. M. Polyakov, A. S. Schwartz
   and Yu. S. Tyupkin, \plb{59}{75}{85}. G. 't Hooft, \prl{37}{76}{8};
   \prd{14}{76}{3432}.
\bibitem{limits} The Aleph Collaboration, \plb{313}{93}{299}.
\bibitem{quasiabelian} E.~S.~Fradkin, ``Proceedings, P.~N.~Lebedev
   Physics Institute" Vol. 29, 123, Consultants Bureau, New York
   (1967).
\bibitem{klimov} V.~V.~Klimov, {\it Sov.~Phys.~JETP}, {\bf 55} (1982) 199.
\bibitem{weldon1} H.~A.~Weldon, \prd{26}{82}{1394}.
\bibitem{weldon2} H.~A.~Weldon, \prd{40}{89}{2410}.
\bibitem{smilga} V.~V.~Lebedev and A.~V.~Smilga, {\it Annals of
   Physics}, {\bf 202}, 229 (1990) and references therein.
\bibitem{damping} E.~Braaten and R.~D.~Pisarsky, \prd{46}{92}{1829}.
\bibitem{Jarlskog} C.~Jarlskog, \prl{55}{85}{1039};
  C. Jarlskog in ``CP Violation,''  ed.~C.~Jarlskog (World
  Scientific, 1989).
\bibitem{ckntop} A.~G.~Cohen, D.~B.~Kaplan and A.~E.~Nelson,
   \npb{373}{92}{453}.
\bibitem{lsvt} L.~McLerran, M.~E.~Shaposhnikov, N.~Turok and
   M.~Voloshin, \plb{256}{91}{451}.
\bibitem{dhss} M.~Dine, P.~Huet, R.~Singleton,~Jr. and L.~Susskind,
   \plb{257}{91}{351}.

\end{thebibliography}
\end{document}